%% file: paper.tex
\documentclass[conference]{IEEEtran}
\IEEEoverridecommandlockouts
\usepackage{cite}
\usepackage{amsmath,amssymb,amsfonts}
\usepackage{algorithmic}
\usepackage{graphicx}
\usepackage{textcomp}
\usepackage[table]{xcolor} 
\newcommand{\greyrule}{\arrayrulecolor{black!30}\midrule\arrayrulecolor{black}}
\definecolor{gold}{RGB}{255,215,0}
\definecolor{silver}{RGB}{192,192,192}
\definecolor{copper}{RGB}{215,155,105}
\newcommand{\cG}{\cellcolor{gold}}
\newcommand{\cS}{\cellcolor{silver}}
\newcommand{\cC}{\cellcolor{copper}}
\definecolor{darkpastelgreen}{rgb}{0.01, 0.75, 0.24}
\def\BibTeX{{\rm B\kern-.05em{\sc i\kern-.025em b}\kern-.08em
    T\kern-.1667em\lower.7ex\hbox{E}\kern-.125emX}}
\usepackage{hyperref}
\hypersetup{%
  colorlinks=true,%
  linkcolor={red!50!black},
  citecolor={blue!65!black},
  urlcolor={blue!80!black},
  bookmarksnumbered=true,
  bookmarksopen=true}
\usepackage{makecell}
\usepackage{booktabs}
\usepackage{multirow}
\usepackage[linesnumbered,ruled,vlined]{algorithm2e}
\usepackage[section]{placeins}
\usepackage{orcidlink}
\newcommand{\sectionname}{Section}

\usepackage{tikz}
\usepackage{subcaption}
\usepackage{threeparttable}
\usepackage{siunitx}
\newcommand*\circledw[1]{\tikz[baseline=(char.base)]{%
            \node[shape=circle,fill=white,draw,inner sep=1pt, minimum size=3mm] (char) {#1};}}
\newcommand\NameOrcid[2]{\hypersetup{hidelinks}\href{https://orcid.org/#2}{#1}}

\begin{document}

\title{SiliconMind-V1: \\Multi-Agent Distillation and Debug-Reasoning Workflows for Verilog Code Generation
\thanks{\scriptsize This work has been submitted to the IEEE for possible publication. Copyright may be transferred without notice, after which this version may no longer be accessible.\\Corresponding authors: \{muchi674, aben20807\}@gmail.com. Relevant resources are available at {https://AS-SiliconMind.github.io/SiliconMind-V1}.}
}


\author{
    \IEEEauthorblockN{%
    \NameOrcid{Mu-Chi Chen}{0009-0007-6013-4122}\IEEEauthorrefmark{1}\IEEEauthorrefmark{2}, %
    \NameOrcid{Yu-Hung Kao}{0009-0002-6991-8795}\IEEEauthorrefmark{2}, %
    \NameOrcid{Po-Hsuan Huang}{0000-0002-7458-9634}\IEEEauthorrefmark{2}, %
    \NameOrcid{Shao-Chun Ho}{0009-0000-7363-2947}\IEEEauthorrefmark{2}, %
    \NameOrcid{Hsiang-Yu Tsou}{0009-0003-1825-0622}\IEEEauthorrefmark{2}, \\%
    \NameOrcid{I-Ting Wu}{0009-0004-0921-1637}\IEEEauthorrefmark{2},
    \NameOrcid{En-Ming Huang}{0000-0003-2196-2834}\IEEEauthorrefmark{2}, %
    \NameOrcid{Yu-Kai Hung}{0009-0007-0310-4883}\IEEEauthorrefmark{2}, %
    \NameOrcid{Wei-Po Hsin}{0009-0000-4015-6083}\IEEEauthorrefmark{2}, %
    \NameOrcid{Cheng Liang}{0009-0009-1532-3332}\IEEEauthorrefmark{2}, \\%
    \NameOrcid{Chia-Heng Tu}{0000-0001-8967-1385}\IEEEauthorrefmark{3}, %
    \NameOrcid{Shih-Hao Hung}{0000-0003-2043-2663}\IEEEauthorrefmark{2}, and %
    \NameOrcid{H. T. Kung}{0000-0002-3348-3788}\IEEEauthorrefmark{4}}\\[-5pt]
    \IEEEauthorblockA{\IEEEauthorrefmark{1}\small Academia Sinica, Taipei, Taiwan}
    \IEEEauthorblockA{\IEEEauthorrefmark{2}\small National Taiwan University, Taipei, Taiwan}
    \IEEEauthorblockA{\IEEEauthorrefmark{3}\small National Cheng Kung University, Tainan, Taiwan}
    \IEEEauthorblockA{\IEEEauthorrefmark{4}\small Harvard University, Cambridge, Massachusetts, USA}\\[-5pt]
    {\url{https://AS-SiliconMind.github.io/SiliconMind-V1}}
}

\maketitle

\begin{abstract}
Large language models (LLMs) have recently emerged as a promising approach for automating Verilog code generation; however, existing methods primarily emphasize syntactic correctness and often rely on commercial models or external verification tools, which introduces concerns regarding cost, data privacy, and limited guarantees of functional correctness. This work proposes a unified multi-agent framework for reasoning-oriented training data generation with integrated testbench-driven verification, enabling locally fine-tuned LLMs, SiliconMind-V1, to iteratively generate, test, and debug Register-Transfer Level (RTL) designs through test-time scaling. Experimental results on representative benchmarks (VerilogEval-v2, RTLLM-v2, and CVDP) demonstrate that the proposed approach outperforms the state-of-the-art QiMeng-CodeV-R1 in functional correctness while using fewer training resources.
\end{abstract}

\begin{IEEEkeywords}
Large Language Models, Verilog Code Generation, Multi-Agent System, Self-Debugging, Reasoning Distillation
\end{IEEEkeywords}

\input{paper_body/introduction}
\input{paper_body/background}
\input{paper_body/architecture}
\input{paper_body/methodology}
\input{paper_body/evaluation}
\input{paper_body/conclusion}

\section*{Acknowledgment}
We acknowledge the financial support from Academia Sinica's SiliconMind Project (AS-IAIA-114-M11). We also thank the National Center for High-Performance Computing (NCHC) for providing computational and storage resources, and Taipei-1 for providing H100 computing resources. In addition, we acknowledge financial support from the National Science and Technology Council.


\FloatBarrier
\bibliographystyle{IEEEtran}
\bibliography{paper}

\end{document}

%% file: paper_body/introduction.tex
\section{Introduction}

Hardware design productivity has become an increasingly critical challenge as modern digital systems continue to grow in scale and complexity. Verilog and SystemVerilog (e.g.,~\cite{verilog_lrm, systemverilog_lrm}) remain the dominant hardware description languages for specifying, verifying, and implementing these systems, yet the development and verification of RTL designs demand substantial domain expertise and manual effort. In recent years, LLMs have shown promising capabilities in code generation and reasoning tasks, motivating growing interest in their application to hardware design automation, particularly for Verilog code generation~\cite{guo2024deepseekcoderlargelanguagemodel, ni-etal-2024-l2ceval}.

At the same time, recent advances in reasoning-oriented LLMs~\cite{DeepSeekR1,openai_gptoss}, test-time scaling techniques~\cite{s1}, and collaborative multi-agent interactions~\cite{hong2024metagpt} suggest a new opportunity for improving hardware code generation.
Early studies demonstrate that LLMs can assist in producing syntactically correct Verilog code and accelerating development workflows. Subsequent work has begun to explore explored fine-tuning strategies, reasoning-oriented distillation, and multi-agent systems to further improve code quality and scalability~\cite{RTLCoder,OriGen,AutoVCoder,VeriSeek,VeriPrefer}. 

However, most existing approaches to Verilog code generation rely heavily on closed-source LLMs and commercial tools during training or verification. This reliance introduces high deployment costs, limits reproducibility, and raises data privacy concerns. Moreover, many of these methods adopt outcome-based reward mechanisms, training models primarily on whether generated code passes syntactic checks or functional tests. These methods assume the correctness of either the generated code or the training data sourced from public repositories. As a result, these outcome-based approaches would overfit to final answers and hence, lead to generalization issue as the code generators. 

Furthermore, although recent advances in LLMs have demonstrated capabilities in reasoning, self-correction, and multi-agent collaboration across complex problem domains, their application to Verilog code generation remains underexplored. Existing studies have yet to systematically investigate how reasoning-oriented training data, testbench-driven functional validation, and collaborative multi-agent inference strategies can be jointly integrated into open-source, small-scale, fine-tuned LLMs. Addressing this gap is particularly important for hardware design.

\begin{figure}[ht]
    \centering
    \includegraphics[width=.86\linewidth]{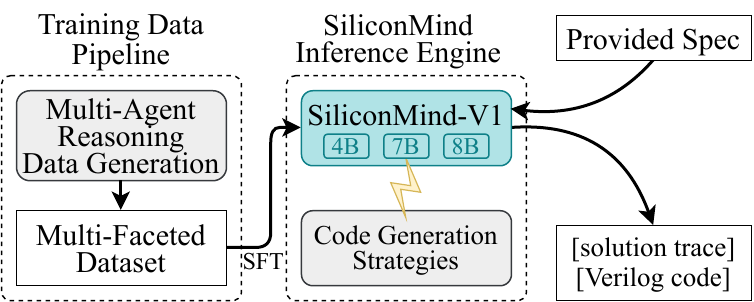}%
    \caption{SiliconMind Framework Overview}%
    \label{fig:framework_overview}%
\end{figure}

The above limitations motivate the need for a unified framework that enables our model, SiliconMind-V1, to reason about, test, and debug Verilog designs while remaining reproducible, cost-efficient, and deployable without reliance on proprietary tools. 
As illustrated in \figurename~\ref{fig:framework_overview}, the framework comprises two core components: a multi-agent pipeline that generates reasoning-rich training data, and a multi-strategy inference engine optimized to exploit these distilled capabilities. Together, these components allow locally fine-tuned LLMs to iteratively generate, test, and debug Verilog code with test-time scaling, completely avoiding dependence on commercial models or external verifiers. The contributions of this work are summarized as follows:

\begin{itemize}
    \item We propose a unified framework that combines multi-agent distillation with test-reasoning workflows for Verilog code generation, where the effectiveness of the inference strategies is enabled by the design of the training data pipeline. To the best of our knowledge, we are the first to propose such a framework, which can be effectively fine-tuned locally to generate, test, and debug Verilog code without external tool use.
    \item We propose a multi-agent data pipeline that automates the generation of reasoning-oriented Verilog design data and testbenches, addressing data scarcity and quality challenges in the hardware domain. As will be demonstrated in \sectionname~\ref{sec:eval_sota}, this reasoning-oriented supervision generalizes better than reward-only alignment. 
    \item We develop a multi-strategy inference engine that guides our distilled LLMs, SiliconMind-V1, to leverage their learned skills for Verilog code generation, testing, and debugging through iterative reasoning and collaboration. Thanks to the reasoning-oriented supervision methodology, the same scaling pattern observed for all the tested open-source, small LLMs can be observed for SiliconMind-V1, which is not the case for outcome-based reward models, as will be shown in \sectionname~\ref{sec:eval_general}.
    \item We conduct extensive experiments using the SiliconMind-V1 series on representative Verilog generation benchmarks, showing that our approach outperforms state-of-the-art methods in functional correctness. Furthermore, when normalized to the performance of the same computing hardware, our approach demonstrates superior efficiency, achieving about 9x speedups in model training, as discussed in \sectionname~\ref{sec:eval_sota}.
\end{itemize}

The remainder of this paper is organized as follows. \sectionname~\ref{sec:background} reviews prior work on LLM-based code generation, provides relevant background, and outlines the motivation for this study. \sectionname~\ref{sec:Architecture} describes the proposed framework architecture, with a focus on the multi-agent data generation pipeline. \sectionname~\ref{sec:Methodology} presents the training methodology and inference strategies. \sectionname~\ref{sec:evaluation} reports experimental results and analysis. Finally, \sectionname~\ref{sec:Conclusion} concludes the paper.

%% file: paper_body/background.tex
\section{Background and Motivation}\label{sec:background}

To provide context for our framework, \sectionname~\ref{sec:background:verilog} introduces Verilog's role in hardware design and verification. This is followed by a review of early fine-tuning approaches for Verilog generation in \sectionname~\ref{sec:background:sft}, while \sectionname~\ref{sec:background:sft_rtl} examines the recent shift toward reasoning-oriented training. Moving to inference-time strategies, \sectionname~\ref{sec:background:multi_agent} discusses training-free multi-agent frameworks that enhance the output of commercial LLMs. These discussions culminate in \sectionname~\ref{sec:background:motivation}, which outlines the primary motivations for this work.

\subsection{Verilog and Testbench}\label{sec:background:verilog}

Verilog and its extension, SystemVerilog, are the primary hardware description languages used to model, simulate, and verify complex digital systems \cite{verilog_lrm, systemverilog_lrm}. Unlike procedural software languages, Verilog explicitly captures concurrency and timing semantics, which are fundamental characteristics of digital logic. By supporting design at the Register-Transfer Level (RTL), Verilog enables efficient architectural modeling and early functional validation. After functional verification, the RTL description is processed by logic synthesis tools, which translate the high-level design into a gate-level netlist mapped to standard cell libraries defined in a target Process Design Kit (PDK) \cite{logic_synthesis}. While logic synthesis and subsequent physical design steps, including placement and routing, are essential for downstream implementation, this work focuses primarily on RTL-level design and verification.

To ensure functional correctness, Verilog designs are typically validated using testbenches, which are non-synthesizable modules created specifically for simulation and verification. A testbench instantiates the design under test (DUT), applies input stimuli through signal assignments or procedural blocks, and observes the corresponding output responses to check whether the design behaves as intended. Testbenches are commonly written in Verilog or SystemVerilog and are kept separate from the RTL to allow flexible control over simulation scenarios without affecting the hardware implementation. Basic testbench components include clock and reset generation, input stimulus drivers, and output monitors that compare observed results against expected values. Through iterative simulation and debugging, designers use testbenches to identify functional errors and validate design correctness before proceeding to synthesis~\cite{LaMeres2024}.

\subsection{Training LLMs for Verilog Generation} \label{sec:background:sft}

The use of LLMs to generate Verilog code has attracted increasing attention in recent years, with both academic research and industrial practice exploring their potential to assist hardware design and development~\cite{RTLCoder,OriGen, AutoVCoder}. Although commercial large language models have demonstrated strong performance in Verilog code generation, concerns about data privacy and high API costs have motivated the development of locally fine-tuned models. A major challenge in this direction is the limited availability of high-quality training datasets. Several prior works have proposed automated data synthesis pipelines to address this issue.

RTLCoder~\cite{RTLCoder} proposed an automated pipeline that synthesizes instruction-code pairs by deriving Verilog design problems from RTL domain keywords or collected Verilog codes and generating the corresponding solution with GPT-3.5. To enhance their model's self-correction capabilities, OriGen~\cite{OriGen} introduced a data pipeline that constructs error-correction examples while procuring instruction-code pairs. However, these two works only guarantee syntactical correctness of their codes. During inference, OriGen even rely on compiler feedback to kickstart the debugging process.

To improve the scale and quality of the training data, AutoVCoder~\cite{AutoVCoder} employs a two-stage fine-tuning strategy. The first stage leverages large amounts of data collected from public GitHub repositories and filtered by a self-trained lightweight code-scorer. Then, the second stage synthesizes instruction-code pairs with ChatGPT-3.5 and functionally verify the code by a testbench also generated by the model. Nonetheless, AutoVCoder does not prepare error-correction training data and depends on Retrieval-Augmented Generation, which allows the model to access external knowledge during inference, to achieve the claimed results.

Recent approaches on fine-tuning LLMs locally for Verilog generation often incorporate Reinforcement Learning with Verifiable Rewards (RLVR). For instance, VeriSeek~\cite{VeriSeek} performs continual pre-training (CPT) on an integrated Verilog and C/C++ corpus followed by Proximal Policy Optimization (PPO) using structure similarity reward, which compares the Abstract Syntax Trees (AST) of the generated code and the reference answer. However, VeriSeek is imited by the inherent quality of its source data: its CPT dataset is entirely unverified, and the instruction-code pairs used during PPO are only syntactically checked.

VeriPrefer~\cite{VeriPrefer} initiates the training process by performing SFT with difficulty filtered and syntactically verified public dataset. Then, the methodology employs Direct Preference Optimization (DPO) to align the model with functionally correct code samples. These samples are generated by the SFT model from the first stage and evaluated using testbenches produced by GPT-4o working cooperatively with Synopsys VCS, an licensed electronic design automation (EDA) tool, that provides coverage report. Aside from relying on a costly EDA tool, VeriPrefer does not hone their model's self correction capabilities nor teach it to reason.

\subsection{Verilog Generation with Trained Reasoning Models} \label{sec:background:sft_rtl}

Recent studies indicate that LLMs can achieve improved performance by thinking before answering. Large Reasoning Models (LRMs), such as DeepSeek-R1~\cite{DeepSeekR1} and gpt-oss-120b~\cite{openai_gptoss}, have demonstrated strong mathematical and coding performance through complex reasoning. As open-source assets, these LRMs can act as teachers to distill reasoning into smaller, specialized LLMs. Illustrating this potential, Muennighoff et al.~\cite{s1} found that SFT on a 32B model using only 1,000 $(problem, reason, answer)$ triplets from DeepSeek-R1 leads to substantial gains in mathematical reasoning.

For Verilog generation, VeriReason~\cite{VeriReason} is the first work to incorporate reasoning-oriented training. First, they jumpstart the base model's reasoning capabilities by performing SFT on ChatGPT-4.1's reasoning traces. Then, they employed Group Relative Policy optimization (GRPO) guided by testbench based functional reward to further refine their model.

QiMeng-CodeV-R1~\cite{QiMeng-CodeV-R1} (CodeV-R1) was previously the state-of-the-art (SOTA) among small-scale LLMs for Verilog design. We define small-scale as models that can be comfortably deployed on gaming GPUs for private hosting. Similar to VeriReason, CodeV-R1 begins training their flagship model by performing SFT with vast amounts of difficulty-filtered $(problem, reason, code)$ data points synthesized via DeepSeek-R1 and V3. Then, they selected the most challenging, high-quality data points for RLVR, which guides the model to prioritize generating functionally correct codes as determined by their automatically (LLM-free) generated testbenches. While both VeriReason and CodeV-R1 reward the model during RLVR when it stumbles upon functionally correct answers and penalize it otherwise, the model is not explicitly learning why its response satisfies the given problem's functional requirements or where it fails to do so. Even subsequent non-reasoning refinements from the same team - such as QiMeng-SALV's~\cite{SALV} signal-level DPO rewards and QiMeng-CRUX's~\cite{CRUX} intermediate specification refinement - have yet to match the performance of the original CodeV-R1.

\subsection{Multi-Agent Inference Systems for Verilog Generation} \label{sec:background:multi_agent}

A multi-agent system (MAS) leverages specialized agents to tackle complex objectives that exceed the capacity of any individual model. By distributing workloads, MAS acts as a mechanism for test-time scaling (TTS), effectively enhancing model performance during inference. When integrated with LLMs, each agent operates as an autonomous entity - reasoning, planning, and communicating in natural language to achieve collective goals.

MetaGPT~\cite{hong2024metagpt} is a pioneering framework for multi-LLM agent software coding that uses standardized operating procedures to structure collaboration. By assigning specialized roles and enforcing standardized outputs, it reduces cascading hallucinations and enables more consistent autonomous handling of complex software tasks.

VerilogCoder~\cite{VerilogCoder} is one of the first works to employ a multi-LLM agent architecture for Verilog code generation. The methodology breaks down the process of implementing a Verilog module from natural language instructions roughly into the following sub-tasks: planning, code generation, and debugging. VerilogCoder's debugging agent depends heavily on feedbacks from the compiler, simulator, and an AST-based waveform tracing tool.

Meanwhile, MAGE~\cite{MAGE} takes a slightly different approach that designs separate agents for testbench generation, code generation, judging and debugging. For early and precise detection of errors, MAGE's testbench and debug agent collaborates to enable a Verilog-state checkpoint mechanism.

\begin{table*}[h]
\centering
\setlength{\tabcolsep}{3.3pt}
\caption{Comparison of LLM-based Verilog Generation Frameworks}
\label{tab:related_work}
\begin{tabular}{lccccccc}
    \toprule
    \textbf{Method} & \textbf{\makecell{Teacher /\\Generator}} & \textbf{\makecell{Student\\Size}} & \textbf{Training} & \textbf{\makecell{Dataset\\Size}} & \textbf{\makecell{Dataset\\Content}} & \textbf{\makecell{Verification\\Level}} & \textbf{\makecell{Test-Time\\Capabilities}} \\
    \midrule
    \multicolumn{7}{l}{\hspace{-10pt}\textit{Fine-tuning Approaches}} \\
    RTLCoder~\cite{RTLCoder} & GPT-3.5 & 6.7B,7B & SFT & 27k & $(p, c)$ & Syntax & Single-Pass \\
    VeriSeek~\cite{VeriSeek} & -- & 6.7B & CPT + RL & $>$109k & $(p, c)$ & Syntax & Single-Pass \\
    AutoVCoder~\cite{AutoVCoder} & GPT-3.5 & 6.7B,7B & SFT & $>$217k & $(p, c)$ & Syntax\& Func & Single-Pass w/ RAG \\
    VeriPrefer~\cite{VeriPrefer} & GPT-4o & 6.7B,7B,14B & SFT + RL & 87k & $(p, c)$ & Syntax \& Func & Single-Pass \\
    CodeV-R1~\cite{QiMeng-CodeV-R1} & DeepSeek-R1, V3 & 7B & SFT + RL & 87k & $(p, r, c, \mathit{tb})$ & Syntax \& Func & Single-Pass \\
    QiMeng-CRUX~\cite{CRUX} & GPT-3.5, DeepSeek-R1 & 7B & SFT + RL & 165k & $(p, X, c)$ & Syntax \& Func & Single-Pass \\
    QiMeng-SALV~\cite{SALV} & GPT-3.5 & 7B & SFT + RL & 135k & $(p, S, c)$ & Syntax \& Func & Single-Pass \\
    \midrule
    \multicolumn{7}{l}{\hspace{-10pt}\textit{Inference-time Approaches}} \\
    VerilogCoder~\cite{VerilogCoder} & GPT-4, Llama3 & -- & -- & -- & -- & Syntax \& Func & \makecell{Agentic$^\dagger$} \\
    MAGE~\cite{MAGE} & Claude-3.5 Sonnet & -- & -- & -- & -- & Syntax \& Func & \makecell{Agentic$^\dagger$} \\
    \midrule
    \multicolumn{7}{l}{\hspace{-10pt}\textit{Hybrid Approaches}} \\
    OriGen~\cite{OriGen} & Claude-3 Haiku & 7B & SFT & 227k & $(p, c, c_\mathit{Err}, \mathit{tf})$ & Syntax & Agentic w/ external tools \\
    \greyrule
    SiliconMind (Ours) & gpt-oss-120b & 4B,7B,8B & SFT & 36k & \makecell{$(p, r, c, \mathit{tb}, \mathit{t\&d})$} & Syntax \& Func & \makecell{Multi-Strategy\\w/o external tool\\w/o benchmark's tb} \\
    \bottomrule
    \addlinespace[1pt]
    \multicolumn{7}{l}{\footnotesize $\dagger$: Agentic LLMs employ external tools and benchmark-provided testbenches for verification.} \\
    \multicolumn{7}{l}{\footnotesize \textbf{Legend:} $p$: problem, $r$: reasoning trace, $c$: code, $c_\mathit{Err}$: erroneous code, $\mathit{tb}$: testbench, $\mathit{t\&d}$: self-testing and debugging traces,}\\
    \multicolumn{7}{l}{\footnotesize \phantom{\textbf{Legend:}} $X$: CRUX artifacts generated by~\cite{CRUX}, $S$: signal-aware data generated by~\cite{SALV}.}
\end{tabular}%
\end{table*}

Note that both VerilogCoder and MAGE's best results were obtained with commercial models such as GPT-4 Turbo and Claude 3.5 Sonnet, which raises data privacy concerns. Moreover, the reliance of VerilogCoder and MAGE on benchmark-provided testbenches compromises their practicality. In real-world scenarios, expecting a user to provide a comprehensive testbench prior to code generation is unrealistic.

\subsection{Motivation}\label{sec:background:motivation}

The reviewed literature shows substantial progress in LLM-based Verilog code generation, particularly through reasoning-oriented models and multi-agent systems, as summarized in Table~\ref{tab:related_work}. However, many existing approaches rely on commercial LLMs, such as GPT and Claude, which raises data privacy concerns and incurs high deployment costs. Methods that only perform syntax checking, depend on external verification tools, or use golden testbench results to assess correctness also limit reliability and practical applicability. Moreover, the reasoning capabilities and test-time scaling behavior of fine-tuned autonomous LLMs remain insufficiently studied for Verilog generation. Prior works often include RLVR stages that reward functional correctness, yet training costs are prohibitively high and models do not explicitly learn from their errors.
To address these issues, a comprehensive framework is needed to automate the generation of high-quality, reasoning-oriented training data and testbenches without relying on commercial LLMs. Such a framework would enable LLMs to generate, test, and debug Verilog code while supporting effective test-time scaling.

%% file: paper_body/architecture.tex
\section{Framework Architecture}\label{sec:Architecture}

\begin{figure*}[ht]
    \centering
    \includegraphics[width=.95\linewidth]{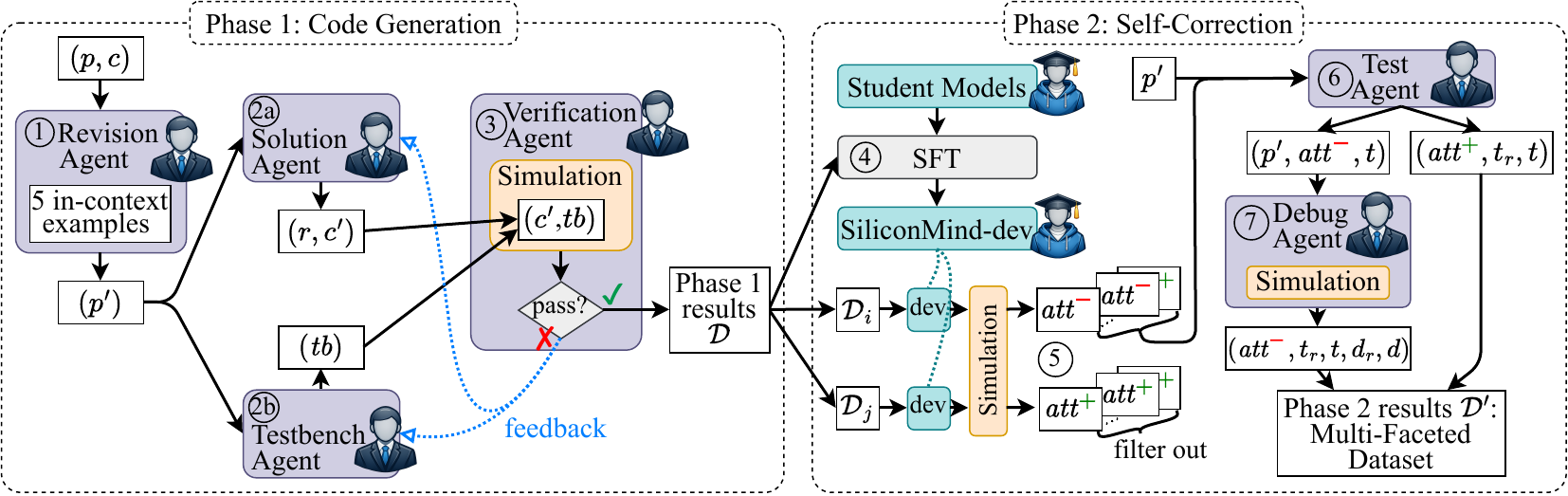}%
    \caption{Training Data Pipeline Overview}%
    \label{fig:data_pipeline}%
\end{figure*}

An overview of the proposed framework is illustrated in \figurename~\ref{fig:framework_overview}. Two major components, the Training Data Pipeline and the SiliconMind Inference Engine, are developed to produce high-quality training data and generate Verilog code based on user specifications. The former addresses the scarcity, quality, and diversity issues common in public Verilog datasets, while the latter enables the fine-tuned LLMs to fully leverage their learned skills to generate the final Verilog code. Details regarding the Data Pipeline and the Inference Engine are elaborated in the following subsections and in \sectionname~\ref{sec:Methodology}, respectively.

The training data pipeline consists of two phases: Training Code Generation (\sectionname~\ref{sec:arch_gen}) and Self-Correction (\sectionname~\ref{sec:arch_correct}), as shown in \figurename~\ref{fig:data_pipeline}. The first phase generates functionally verified Verilog code for general training. The second phase analyzes the limitations of the model trained in the first phase and enriches existing data with targeted testing and debugging curriculum.

The proposed model series, SiliconMind-V1, is subsequently trained on the output of this pipeline. Furthermore, three TTS code generation strategies are integrated by using the SiliconMind Inference Engine. These strategies enable the LLMs (SiliconMind-V1) to effectively utilize their learned knowledge to generate code that consistently satisfies user specifications. The following describes the two components of the Training Data Pipeline and explain their interactions. It is worth noting that for all of our pipeline's generative tasks, we picked the best open-source LLM our hardware resource could afford, gpt-oss-120b~\cite{openai_gptoss} as the teacher model.

\subsection{Training Code Generation} \label{sec:arch_gen}

In the \textit{Training Code Generation} phase, the pipeline employs four specialized agents to create functionally verified training data for downstream Verilog code generation. It is important to note that in order to provide high-quality training data, the mission of this phase is to generate refined problem statements and corresponding Verilog codes, reasoning data, and relevant testbenches. The following describes steps taken by multiple agents to achieve this goal.

\emph{\circledw{1} Revision Agent}. This  agent  takes Verilog design problems ($p$) and their solution codes ($c$) as inputs, as illustrated in \figurename~\ref{fig:data_pipeline}. It refines $p$ into $p^\prime$, ensuring that module names, port lists, and design behaviors are explicitly defined to reduce false negatives during downstream functional verification. It is worth noting that $p$ is refined by the filtered solutions to ensure that $p^\prime$ accurately reflects the functionality implemented in those filtered $c$. Also note that the refined problem $p^\prime$ is then used in downstream agents for generating the corresponding solution $c^\prime$ and testbench $tb$.

The $(p, c)$ pairs are collected from public sources including DeepCircuitX~\cite{DeepCircuitX}, PyraNet~\cite{PyraNet}, RTLCoder~\cite{RTLCoder}, VeriThought~\cite{VeriThoughts}, and Verilog\_Github~\cite{verilog_github}. Based on our analysis, many $c$ contain syntactical errors, and they could also fail to satisfy $p$'s functional requirements even if they are syntactically correct. Under this circumstance, the revision agent filters out those data points in $c$ that cannot be compiled using an open-source EDA tool Icarus Verilog, and performs the functional correctness check for the remaining $c$. Particularly, the 5-shot prompt technique ~\cite{QiMeng-CodeV-R1} is employed to examine each $p^\prime$ such that the $p^\prime$ matches the corresponding solution in $c$. 
The prompt gives five examples that breakdown $p^\prime$ generation from $c^\prime$ into two steps: 1) describing the behavior of $c^\prime$ and 2) deriving the formal problem statement $p^\prime$ from said description. The refined problems $p^\prime$ are then used in the Solution and Testbench Agents for generating the corresponding solution $c^\prime$ and testbench $tb$. Note that the open source dataset would contain erroneous solution codes. Instead of fixing the errors within the solution code, this work uses the open source solution code to provide higher-quality, more precise problem descriptions ($p^\prime$), which is used to generate the required solutions and test programs.

\emph{\circledw{2a} Solution Agent}. Given the refined problem ($p^\prime$),  the solution agent tries to reason deeply ($r$) before providing the final answer ($c^\prime$). It is worth noting that we chose not to take $p^\prime$ and $c$ as inputs and ask for the reasoning data that connects the two. This decision is made based on the observation that LLMs struggle to generate the thought process with the given problem and solution pair. The resulting $r$ and $c^\prime$ are then sent to the verification agent for functional verification. If the verification agent detects an error in the first attempt, the solution gent is granted one retry. We avoided asking the solution agent to debug the initial attempt with external tool feedback because we want $r$ to be solely about solving a problem from scratch.

\emph{\circledw{2b} Testbench Agent}. This agent is prompted with $p^\prime$ and instructed to produce a testbench ($tb$) that meets several criteria: it must include representative test cases and meaningful error messages, remain compatible with Icarus Verilog~\cite{iverilog}, and be compilable with an external design under test file. If the verification agent detects any issues with $tb$, it provides an error report for debugging.

\emph{\circledw{3} Verification Agent}. As the referee, the verification agent collects $p^\prime$ from the revision agent, $c^\prime$ from the solution agent, and $tb$ from the testbench agent. Then, it simulates the code with the testbench for functional verification. If the simulation passes, the verification agent adds $(p^\prime,\ r,\ c^\prime,\ tb)$ to the training dataset. Otherwise, the verification agent refers to the tool's response along with $p^\prime$ to determine, in the format of an error report, whether the solution code or the testbench is at fault. Depending on the diagnosis result, the verification agent either asks the solution agent to provide a new $c^\prime$ or sends the error report to the testbench agent to debug $tb$. Finally, if the updated $(c^\prime,\ tb)$ continues to fail during simulation, the data point (denoted by $p^\prime$) is discarded.

As a result, in the \textit{Training Code Generation} phase, we obtained 36k $(p^\prime,\ r,\ c^\prime,\ tb)$ tuples from publicly sourced $(p, c)$ pairs, denoted as $\mathcal{D}$. Here, $p^\prime$ is clearly defined, and $c^\prime$ is functionally verified by $tb$.

\subsection{Self-Correction} \label{sec:arch_correct}

In the \textit{Self-Correction} phase, the pipeline leverages the model trained in the \textit{Training Code Generation} phase to identify its weaknesses and augment existing data points with tailored testing and debugging curriculum from two additional agents.

\emph{\circledw{4} Internal SFT}. The training workflow trains a base model on $\mathcal{D}$, the 36k $(p^\prime,\ r,\ c^\prime)$ tuples by the verification agent. The resulting model, named \textit{SiliconMind-dev}, learns to reason before generating the final code for the given Verilog design problem.

\emph{\circledw{5}} The pipeline prompts \textit{SiliconMind-dev} with each data tuple in $\mathcal{D}$, asking it to generate a new solution code for each $p^\prime$. These generated solution codes are tested against the corresponding $tb$ for functional correctness via simulations.  Solution codes that pass the $tb$ in simulations  are labeled as $att^{+}$, while those that fail are denoted as $att^{-}$. For those $p^\prime$ where \textit{SiliconMind-dev} fails at least once, they are selected for further processing in the subsequent steps.

\emph{\circledw{6} Test Agent}. The test agent collects problems ($p^\prime$) that \textit{SiliconMind-dev} sometimes gets wrong (i.e., $\geq 1$ of the model's attempts are marked wrong by $tb$). Next, given $(p^\prime,\ att^{\pm})$, the test agent reasons deeply ($t_r$) before writing a test report ($t$) about what is right/wrong about $att^{+}$/$att^{-}$. If both $t$ and $tb$ agrees that $att^{+}$ is correct, $(att^{+}, t_r, t)$ will be added to the $p^\prime$ denoted data point, as part of the final training dataset, $\mathit{D}^\prime$.

For each selected $p^\prime$, the test agent removes exact duplicate $att^{\pm}$ samples and balances the retained $att^{+}$ and $att^{-}$ instances. Instead of providing the testbench $tb$, we instruct the test agent to derive a couple representative test cases from $p^\prime$ and reason about the behavior of $att^{\pm}$ under each case. This process encourages $t_r$ to approximate a mental walkthrough of $att^{\pm}$ without relying on external tools.

\emph{\circledw{7} Debug Agent}. This agent is responsible to provide debugging solutions for faulty results obtained in the Step \circledw{5}, specifically those containing errors in $t$ and $tb$ relative to $p^\prime$. To achieve this, the pipeline prompts the debug agent with $(p^\prime,\ att^{-},\ t)$ to perform deep reasoning ($d_r$) before generating a corrected solution ($d$).

Next, if $d$ passes $tb$ after the simulation, $(att^{-}, t_r, t, d_r, d)$ is added to the $p^\prime$, and included in $\mathit{D}^\prime$.
Otherwise, if the first debug attempt ($d_{old}$) fails $tb$, the debug agent initiates a second iteration by asking the test agent for a report on $d$. This results in two versions of $(t_r, t, d_r, d)$: one based on $att^{-}$, and another on $d_{old}$. To keep training sequence length manageable, we only append the version based on $d_{old}$ to $\mathit{D}^\prime$ if the subsequent attempt $d_{new}$ passes $tb$.

The \textit{Self-Correction} phase augments the dataset $\mathit{D}$ with $(att^{+}, t_r, t)$ and $(att^{-}, t_r, t, d_r, d)$ data points, representing problems where \textit{SiliconMind-dev} exhibited weaknesses. In these cases, both $t$ and $d$ are functionally verified using the corresponding testbench $tb$.

Upon completion of the Training Data Pipeline, a multi-faceted training dataset $\mathit{D}^\prime$ is produced. This pipeline then fine-tunes \textit{SiliconMind-dev} on the newly generated data, resulting in a model referred to as \textit{SiliconMind-V1}, as illustrated in \figurename~\ref{fig:framework_overview}. Through this process, the model learns to test and debug its own generated Verilog code. The next section presents the training methodology and the multi-strategy inference engine that guides \textit{SiliconMind-V1} during Verilog design tasks.

%% file: paper_body/methodology.tex
\section{Model Training Methodology}\label{sec:Methodology}

The proposed methodology for training large language models and guiding them during inference to generate, test, and debug Verilog code is introduced in this section. \sectionname~\ref{sec:method:internalsft} describes the SFT process used to produce the \textit{SiliconMind-dev} models (Step \circledw{4} in \sectionname~\ref{sec:arch_correct}), while \sectionname~\ref{sec:method:tailoredsft} details the tailored SFT procedure for creating the \textit{SiliconMind-V1} models (the SFT process illustrated in \figurename~\ref{fig:framework_overview}). After training, \sectionname~\ref{sec:method_tts} presents the multi-strategy inference engine that guides the trained models in performing Verilog design tasks.

To demonstrate the generalizability of our proposed framework, four LLMs are selected as base models in the model training. They are \textit{Qwen2.5-Coder-7B-Instruct}~\cite{qwen_coder}, \textit{Qwen3-4B-Thinking-2507} and \textit{Qwen3-8B}~\cite{qwen3}, and \textit{Olmo-3-7B-Think}~\cite{groeneveld-etal-2024-olmo}. The first three are selected for their competitive performance in code generation and reasoning tasks, while the last is chosen for its fully open-sourced nature, despite its limited Verilog design capabilities.

\subsection{SFT for SiliconMind-dev Models}\label{sec:method:internalsft}

The four base models are initially fine-tuned on the 36k dataset from the \emph{Training Code Generation} phase to produce the \emph{SiliconMind-dev} models, as shown in Step \circledw{4} of \figurename~\ref{fig:data_pipeline}. Specifically, for a given Verilog design problem $p^\prime$, the models are trained to reason about the solution before generating the final code $c^\prime$ along with its corresponding reasoning trace $r$:

\begin{gather*}
\text{given }p^\prime \underset{\text{output}}{\rightarrow} (r,\ c^\prime)
\end{gather*}

The resulting models are collectively referred to as \emph{SiliconMind-dev}. We first conduct a preliminary evaluation to select representative development models from different model families for the \emph{Self-Correction} phase. Specifically, we choose a \emph{Qwen2.5-Coder-7B-Instruct}-based model from the Qwen family and an \emph{Olmo-3-7B-Think}-based model from the Olmo family. As shown in Table~\ref{tab:data2_res}, these selected \emph{dev} models participate in the \emph{Self-Correction} phase, which augments 6.8k and 6.2k original data points, respectively, with tailored testing and debugging curriculum.
Within the Qwen family, the \emph{Qwen2.5-Coder-7B-Instruct}-based dev model is deliberately chosen because it underperforms compared to its stronger counterparts, such as \emph{Qwen3-4B-Thinking-2507} and \emph{Qwen3-8B}. This choice is motivated by the observation that weaker models offer greater room for improvement and can thus provide clearer evidence of the effectiveness of our approach. By applying the tailored SFT process to a less capable baseline, we more clearly demonstrate how iterative reasoning and self-correction enhance testing and debugging abilities, leading to improved functional correctness.

The details of the SFT process are as follows. Full-parameter SFT is employed to train the models, with the objective of predicting the next token in a sequence (input + output) given all preceding tokens. Formally, during SFT, the model minimizes the average negative log-likelihood loss:  
$$L = -\sum_{t=1}^{T} \log P_{\theta}(x_t \mid x_{<t})$$  
where $x_t$ is the token at timestep $t$ and $P_{\theta}$ is the probability assigned by the model to that token. Table~\ref{tab:train_settings} summarizes the configuration of our SFT experiments.

\begin{table}[ht]
    \centering
    \caption{Supervision Experiment Settings}\label{tab:train_settings}
    \begin{tabular}{l|l}
    \toprule
    \textbf{Parameter} & \textbf{Value}      \\ \midrule
    Completion only loss    & True                \\
    Gradient checkpointing  & True                \\
    Packing                 & True                \\
    Mixed precision         & BF16$^*$           \\
    Epochs                  & 6                   \\
    Effective batch size    & 32                  \\
    Max sequence length     & 30K tokens          \\
    Learning rate scheduler & cosine              \\
    Learning rate           & 2e-5                \\
    Optimizer               & AdamW               \\
    Warmup ratio            & 0.03                \\ \bottomrule
    \addlinespace[1pt]
    \multicolumn{2}{c}{\makecell[l]{$*$: Weights are updated in FP32, while forward\\\phantom{$*$:} and backward passes use BF16.}}
    \end{tabular}
\end{table}

\subsection{SFT for SiliconMind-V1 Models}\label{sec:method:tailoredsft}

\begin{table}[]
\centering
\caption{Number of data points processed by the \textit{Self-Correction} phase for each \textit{SiliconMind-dev} model}\label{tab:data2_res}
\begin{tabular}{l|c}
\toprule
\textbf{\emph{SiliconMind-dev} Model} & \textbf{\makecell{\emph{Self-Correction} Phase\\ \# data points}} \\ \midrule
Qwen2.5-Coder-7B-Instruct & 6.8K \\ 
Olmo-3-7B-Think   & 6.2K \\ \bottomrule
\end{tabular}
\end{table}

Given the augmented data $D^\prime$ from the \emph{Self-Correction} phase, the SFT process is further tailored to train the \textit{SiliconMind-dev} models for improving self-testing and debugging capabilities. Two tasks are designed to achieve this goal.
\begin{enumerate}
    \item Given a Verilog design problem and an attempted solution, think about representative test scenarios and how the provided code behaves under them before organizing a test report. $$\text{given }(p^\prime, att^{\pm}) \underset{\text{output}}{\rightarrow} (t_r,\ t)\text{,}$$
    \item Given a Verilog design problem, an attempted solution, and a test report, think about how to leverage the test report for debugging before providing the final corrected code. $$\text{given }(p^\prime,\ att^{-},\ t) \underset{\text{output}}{\rightarrow} (d_r,\ d)$$
\end{enumerate}
The resulting models are named \emph{SiliconMind-V1}. Through this training, \emph{SiliconMind-V1} learns to generate test reports and debug Verilog code with them, all without external tool use.

\subsection{Inference Strategies} \label{sec:method_tts}
\begin{figure}[ht]
    \centering
    \includegraphics[width=.9\linewidth]{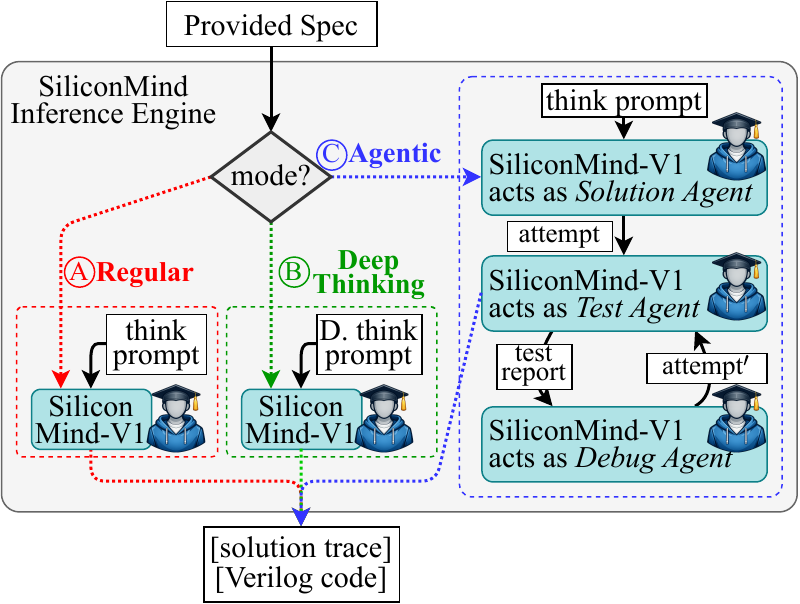}%
    \caption{SiliconMind Inference Engine}%
    \label{fig:workflow_overview}%
\end{figure}

After training our models to generate, test, and debug Verilog code, we devised three inference strategies, including \textit{Regular}, \textit{Deep Thinking}, and \textit{Agentic}, to realize their full potential when addressing Verilog design problems (see \figurename~\ref{fig:workflow_overview}). These strategies enable users to scale the model's reasoning effort prior to generating a final solution.

\emph{\circledw{A} Regular Strategy}. In the \textit{Regular} strategy, when given a Verilog design problem, we prepend a system prompt that asks the model to think before providing the solution code. Here, we are trusting that the model would leverage its newly acquired skillsets to tackle the provided tasks.

\emph{\circledw{B} Deep Thinking Strategy}. We include explicit instructions in the system prompt that asks the model to solve the provided Verilog design problem by coming up with an initial solution, testing it, and debugging it if necessary in its reasoning trace. While this softly coerces the model to leverage everything in its toolbox, there is no guarantee that the model could comfortably extend its reasoning trace to do everything at once without drifting off.

\emph{\circledw{C} Agentic Strategy}. We programmatically separate generating an initial solution, testing, and debugging into three requests. As illustrated in the right of \figurename~\ref{fig:workflow_overview}, this allows repeated testing and debugging when solving a Verilog design problem. Note that the number of interactions between the test and debug agent can either be a manually defined budget or infinity, which lets the model continues to refine its answer until it is satisfied.

These three strategies were designed with cost-performance trade-off in mind. Ideally, \textit{Deep Thinking} should warrant a higher token budget (i.e., longer response length \& latency) than the \textit{Regular} strategy but offer better accuracy. The \textit{Agentic} strategy should be the most costly but offers the best accuracy.

%% file: paper_body/evaluation.tex
\section{Evaluation}\label{sec:evaluation}

In this section, we provide a comprehensive evaluation of the \textit{SiliconMind-V1} model family. We begin by detailing our experimental setup, evaluation metrics, and the benchmarks used in \sectionname~\ref{sec:eval_setup}. We then compare our models against SOTA domain-specific models and general-purpose foundation models in Sections~\ref{sec:eval_sota} and \ref{sec:eval_general}, highlighting our superior performance achieved with significantly fewer training resources. Furthermore, we conduct an ablation study in \sectionname~\ref{sec:eval_ablation} to quantify the impact of our training stages and inference strategies. Finally, we analyze the benefits of curriculum tailoring in \sectionname~\ref{sec:eval_tailored} and discuss the trade-offs between inference token cost and accuracy in \sectionname~\ref{sec:eval_cp}.

\begin{table*}[h]
\centering
\caption{Main Results: Pass@k=1,3,5 Performance (\%)}
\label{tab:eval_general}
\begin{tabular}{ll *{12}{w{c}{14pt}}}
\toprule
 & & \multicolumn{3}{c}{\textbf{RTLLM-v2}} & \multicolumn{3}{c}{\textbf{VerilogEval-v2}} & \multicolumn{3}{c}{\textbf{VerilogEval-v2-NTU}} & \multicolumn{3}{c}{\textbf{CVDP-cid02\&03}} \\
\cmidrule(lr){3-5} \cmidrule(lr){6-8} \cmidrule(lr){9-11} \cmidrule(lr){12-14}
\textbf{Model Name} & \multicolumn{1}{c}{\textbf{Base Model}} & \textbf{p@1} & \textbf{p@3} & \textbf{p@5} & \textbf{p@1} & \textbf{p@3} & \textbf{p@5} & \textbf{p@1} & \textbf{p@3} & \textbf{p@5} & \textbf{p@1} & \textbf{p@3} & \textbf{p@5} \\
\midrule
\multicolumn{14}{l}{\hspace{-10pt}\textit{Foundation Models:}}\\
DeepSeek-R1-0528      & \multicolumn{1}{c}{--} & 68.7 & 75.7 & 77.3 & 80.9 & 88.1 & 90.2 & 86.4 & 93.4 & 95.5 & 25.6 & 32.8 & 35.5 \\
gpt-oss-120b (high)   & \multicolumn{1}{c}{--} & 70.0 & 75.8 & 78.2 & 83.2 & 89.7 & 91.2 & 87.9 & 94.2 & 95.6 & 27.6 & 35.1 & 37.7 \\
Qwen3-32B             & \multicolumn{1}{c}{--} & 55.4	& 67.5 & 70.7 & 70.3 & 80.7 & 83.2 & 76.3 & 86.1 & 88.6 & 12.8 & 20.4 & 23.9 \\
Qwen3-14B             & \multicolumn{1}{c}{--} & 50.0	& 61.8 & 66.5 & 64.2 & 74.4 & 77.9 & 69.5 & 80.1 & 82.9 & 12.9 & 18.7 & 21.6 \\
\greyrule
Qwen2.5-C-7B-I        & \multicolumn{1}{c}{--} & 29.3 & 42.6 & 48.6 & 31.5 & 45.4 & 50.8 & 33.6 & 48.0 & 53.7 & 7.3  & 12.7 & 15.3 \\
Qwen3-4B-T-2507       & \multicolumn{1}{c}{--} & 36.4 & 46.7 & 50.9 & 48.2 & 56.5 & 59.7 & 52.5 & 62.1 & 65.4 & 12.4 & 17.3 & 19.4 \\
Qwen3-8B              & \multicolumn{1}{c}{--} & 40.2 & 55.2 & 61.1 & 53.7 & 65.1 & 69.1 & 57.4 & 70.0 & 74.3 & 11.9 & 16.9 & 19.4 \\
Olmo-3-7B-Think       & \multicolumn{1}{c}{--} & 10.4 & 20.0 & 24.8 & 7.8  & 18.5 & 25.8 & \phantom{0}8.9  & 20.7 & 28.3 & \phantom{0}1.2  & \phantom{0}3.0  & \phantom{0}4.2  \\
\midrule
\multicolumn{14}{l}{\hspace{-10pt}\textit{Fine-tuned Models:}}\\
CodeV-R1-7B-Distill   & Qwen2.5-C-7B-I & 58.5 & 68.6 & 72.5 & 66.4 & 75.5 & 78.5 & 69.6 & 78.8 & 81.7 & 19.0 & 27.5 & 31.0 \\
CodeV-R1-7B           & Qwen2.5-C-7B-I & \cC\textbf{66.1} & \cS\textbf{73.2} & \cG\textbf{75.5} & \textbf{69.7} & \textbf{76.5} & 78.7 & 73.2 & 81.0 & \textbf{83.6} & 21.3 & 28.0 & 30.8 \\
\greyrule
\multirow{4}{*}{SiliconMind-V1} & Qwen2.5-C-7B-I    & 63.8     & 71.9     & 74.0     & \textbf{69.7} & \textbf{76.5} & \textbf{78.8} & \textbf{73.9} & \textbf{81.2} & \textbf{83.6} & \cC\textbf{22.3} & \cC\textbf{30.1} & \cC\textbf{32.7} \\
                                & Qwen3-4B-T-2507   & \cG 67.9 & \cG 74.0 & \cS 75.3 & \cS 76.4      & \cS 82.2      & \cS 83.9      & \cG 82.0      & \cG 88.1      & \cS 89.6      & \cS 23.5         & \cS 30.7         & \cS 33.4 \\
                                & Qwen3-8B          & \cS 66.6 & \cC 73.1 & \cC 74.9 & \cG 76.5      & \cG 82.5      & \cG 84.7      & \cS 81.0      & \cS 87.7      & \cG 89.8      & \cG 24.0         & \cG 31.9         & \cG 35.2 \\
                                & Olmo-3-7B-Think   & 63.3     & 70.8     & 72.6     & \cC 73.5      & \cC 79.4      & \cC 81.1      & \cC 79.5      & \cC 86.7      & \cC 88.6      & 21.2             & 29.1             & 32.0 \\
\bottomrule
\addlinespace[1pt]
\multicolumn{14}{l}{\footnotesize \textbf{Note:} Bold values denote the better-performing model between CodeV-R1 and ours using the same base model.} \\
\multicolumn{14}{l}{\footnotesize Colors denote rankings among specialized models: \colorbox{gold}{first}, \colorbox{silver}{second}, and \colorbox{copper}{third}.} \\
\multicolumn{14}{l}{\footnotesize For brevity, we refer to \textit{Qwen2.5-Coder-7b-Instruct} as \textit{Qwen2.5-C-7B-I} and \textit{Qwen3-4B-Thinking-2507} as \textit{Qwen3-4B-T-2507}.} \\
\end{tabular}
\end{table*}

\subsection{Experimental Setup}\label{sec:eval_setup}

To enable efficient inference with our models and baseline targets for comparison, we implemented a custom inference engine based on \textit{vLLM} version 0.11.2~\cite{kwon2023efficient}. The engine supports three inference strategies and uses the following settings: temperature${=}1.0$, repetition\_penalty${=}1.0$, top\_k${=}{-}1$, and top\_p${=}1.0$ or $0.9$. All reported inference results were obtained in our own benchmark environment on a single NVIDIA DGX H100 node, which is equipped with eight H100 SXM GPUs with a full NVLink interconnect. These results, including those in Table~\ref{tab:eval_general}, were not taken from existing literature.

In the following section, model performance is measured by $pass@k$~\cite{Codex}, the probability that at least one of the k generated solutions is correct for a given problem. Formally, the metric is defined as:
$$pass@k = \mathbb{E} \left[ 1 - \frac{\binom{n-c}{k}}{\binom{n}{k}} \right]$$
\begin{itemize}
    \item $\mathbb{E}$: The average over all problems in the benchmark.
    \item $n$: The total number of samples generated per problem.
    \item $c$: The number of correct samples.
    \item $k$: The number of evaluated samples ($k \le n$).
\end{itemize}
We measured up to $k=1, 3, 5$, $n=20$ to reduce variance and better grasp the model's potential when multiple attempts are allowed. In all following tables and figures, we multiplied $pass@k$ by 100 for better readability.

Our models were mainly evaluated on three benchmarks: 50 problems from RTLLM-v2~\cite{RTLLMv2}, 156 code generation problems from VerilogEval-v2~\cite{VerilogEvalV2}, and 172 code completion and generation problems from CVDP~\cite{CVDP}. Note that, for CVDP, we only picked the most relevant categories, cid02 \& 03, to our work. We also identified several issues in VerilogEval-v2, including inaccurate or ambiguous problem descriptions, unsynthesizable Verilog syntax, and logical inconsistencies in the reference answer. Therefore, with the help of IC domain experts, we created the VerilogEval-v2-NTU benchmark by resolving issues we found in the original dataset.

\subsection{Comparison with previous SOTA} \label{sec:eval_sota}

\begin{table}[h]
\centering
\caption{Cosine Similarity between Centroids of Training Dataset and Benchmark}
\label{tab:eval_contaminate}
\renewcommand{\arraystretch}{1.2} 
\begin{tabular}{lcc}
\toprule
\textbf{Training Dataset} & \textbf{RTLLM-v2} & \textbf{VerilogEval-v2-NTU} \\ 
\midrule
CodeV-R1 (87k) & 0.95 & 0.82 \\
\textbf{Ours (36k)}   & 0.92 & 0.86 \\ 
\bottomrule
\end{tabular}
\end{table}

CodeV-R1~\cite{QiMeng-CodeV-R1}'s 7B model, fine-tuned from \textit{Qwen2.5-Coder-7B-Instruct}, was the previous SOTA small-scale LLM for Verilog code generation. Their SFT stage uses 87k data points, and the 3.1k most challenging and high quality data points were further used for RLVR. In total, CodeV-R1 expended 2,656 A100-80G GPU hours for training.

On the other hand, we built the \textit{SiliconMind-V1} models with a significantly more lightweight approach, only using 36k functionally verified data points and 92 H100-SXM GPU hours of SFT (to train the \textit{Qwen2.5-Coder-7B-Instruct} variant).
When normalized to account for the 3.2x performance leap from the A100 to the H100 (using BF16 Tensor Core), the training time for SiliconMind-V1 could equate to 294.4 A100 GPU hours. This represents a 9x speedup compared to the CodeV-R1 approach. 

Table~\ref{tab:eval_general} presents our models' performance on different benchmarks when using the \textit{Agentic} inference strategy. For budget control, we limit the number of interactions between the test \& debug agent to three. As bold faced on the table, our \textit{Qwen2.5-Coder-7B-Instruct} based \textit{SiliconMind-V1} model outperforms CodeV-R1 on VerilogEval-v2-NTU and CVDP-cid02\&03, matches it on VerilogEval-v2, and trails slightly on RTLLM-v2.

CodeV-R1 acknowledges that its advantage on RTLLM-v2 is primarily attributed to RLVR training on data points more closely aligned with the benchmark. We further confirmed this by using \textit{jina-code-embeddings-1.5b}~\cite{JINA} to transform solution codes in the training datasets and the benchmarks to mathematical vectors, computing each set of vectors' centroids, and comparing the centroids' cosine similarities. As shown in Table~\ref{tab:eval_contaminate}, CodeV-R1's dataset exhibits an anomalously high similarity of 0.95 with RTLLM-v2, which drops sharply to 0.82 on VerilogEval-v2-NTU (a 0.13 delta). This discrepancy suggests that CodeV-R1's dataset is heavily skewed toward the RTLLM-v2 distribution. In contrast, our dataset maintains a more consistent alignment across both benchmarks (0.92 to 0.86, a 0.06 delta), indicating a training distribution that is robust and generalizable. Note that we could not perform the same analysis with CVDP-cid02\&03 since the benchmark only uses testbenches and not reference answers.

\begin{table}[ht]
\centering
\caption{Ablation Study: Pass@1 Performance Delta from Framework Progression}
\label{tab:eval_ablation}
\begin{tabular}{l ll ccc}
\toprule
\multirow{4}{*}{\textbf{Base Model}} & \multicolumn{2}{c}{\textbf{Progression}} & \multicolumn{3}{c}{\textbf{Pass@1 $\Delta$ (\%)}} \\
\cmidrule(lr){2-3} \cmidrule(l){4-6}
& \multicolumn{1}{c}{\textbf{From}} & \multicolumn{1}{c}{\textbf{To}} & {\textbf{\makecell{RT}}} & {\textbf{\makecell{VE}}} & {\textbf{\makecell{CV}}} \\
\midrule
\multirow{4}{*}{\makecell{Qwen2.5\\-C-7B-I}} 
    & base        & dev-regular       & 26.7 & 29.2 & 12.5 \\
    & dev-regular & V1-regular        & \phantom{0}5.3  & \phantom{0}7.7  & \phantom{0}2.1  \\
    & V1-regular  & V1-D.Thinking     & \phantom{0}0.8  & \phantom{0}1.0  & \phantom{0}0.3  \\
    & V1-regular  & V1-Agentic & \phantom{0}2.5  & \phantom{0}3.4  & \phantom{0}0.4  \\
\midrule
\multirow{4}{*}{\makecell{Qwen3-4B\\-T-2507}}   
    & base        & dev-regular & 21.6 & 21.9 & \phantom{-}9.7  \\
    & dev-regular & V1-regular & \phantom{0}6.3  & \phantom{0}4.2  & \phantom{-}0.4  \\
    & V1-regular  & V1-D.Thinking  & \phantom{0}0.6  & \phantom{0}0.5  & -0.4 \\
    & V1-regular  & V1-Agentic & \phantom{0}3.6  & \phantom{0}3.4  & \phantom{-}1.0  \\
\midrule
\multirow{4}{*}{Qwen3-8B}   
    & base        & dev-regular & 19.8 & 18.2 & 10.2 \\
    & dev-regular & V1-regular & \phantom{0}4.8  & \phantom{0}3.0  & \phantom{0}1.3  \\
    & V1-regular  & V1-D.Thinking  & \phantom{0}1.1  & \phantom{0}0.6  & \phantom{0}0.6  \\
    & V1-regular  & V1-Agentic & \phantom{0}1.8  & \phantom{0}2.4  & \phantom{0}0.6  \\
\midrule
\multirow{4}{*}{\makecell{Olmo-3-7B\\-Think}}   
    & base        & dev-regular & 35.6 & 59.7 & 14.5 \\
    & dev-regular & V1-regular & 11.2 & \phantom{0}6.1  & \phantom{0}3.1  \\
    & V1-regular  & V1-D.Thinking  & \phantom{0}2.6  & \phantom{0}0.8  & \phantom{0}1.6  \\
    & V1-regular  & V1-Agentic & \phantom{0}6.1  & \phantom{0}4.8  & \phantom{0}2.4  \\
\bottomrule
\addlinespace[1pt]
\multicolumn{6}{l}{\footnotesize RT=RTLLM-v2, VE=VerilogEval-v2-NTU, CV=CVDP-cid02\&03.} \\
\multicolumn{6}{l}{\footnotesize \textbf{Note:} For the \textit{Agentic} strategy, we limit the number of test/debug} \\
\multicolumn{6}{l}{\footnotesize \phantom{\textbf{Note:}} agent interactions to three.}
\end{tabular}
\end{table}

\begin{figure*}[h]
    \centering
    \includegraphics[width=.98\linewidth]{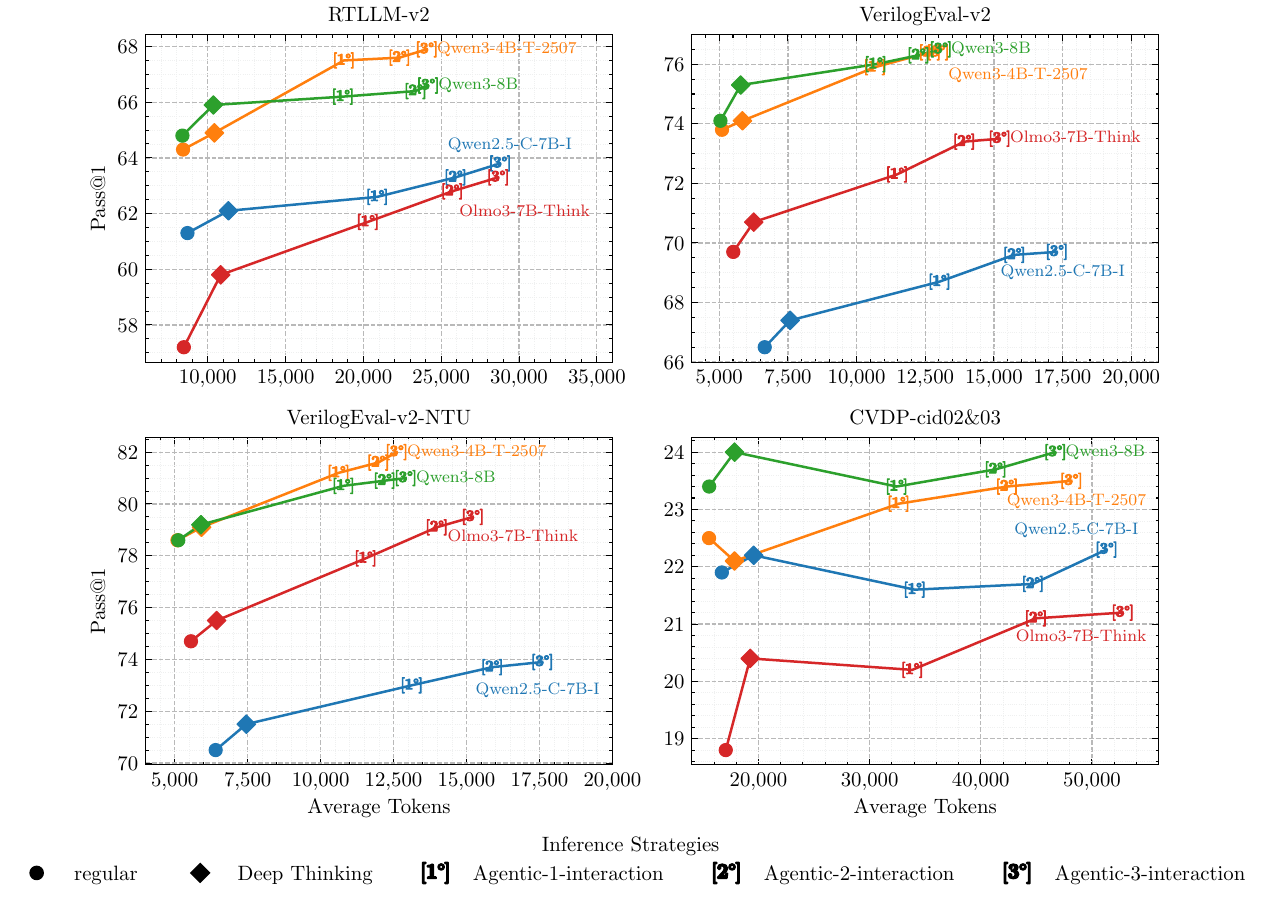}
    \caption{Pass@1 Performance (\%) vs. Average Token Cost Trade-off}
    \label{fig:eval_costperf}
\end{figure*}

\subsection{Generalizability} \label{sec:eval_general}

Table~\ref{tab:eval_general} further highlights the competitive performance of other \textit{SiliconMind-V1} variants. Against all odds discussed in \ref{sec:eval_sota}, our \textit{Qwen3-4B-Thinking-2507} and \textit{Qwen3-8B} variants exceed \textit{CodeV-R1}'s \textit{pass@1} accuracy on RTLLM-v2 while maintaining comparable \textit{pass@3} and \textit{pass@5} metrics. Notably, on the VerilogEval-v2, VerilogEval-v2-NTU, and CVDP-cid02\&03 benchmarks, these models consistently surpass \textit{CodeV-R1} by margins of 2.2-8.8\%.

The efficacy of our methods is further demonstrated when applied to \textit{Olmo-3-7B-Think}, a model with barely any Verilog design capability. The resulting \textit{SiliconMind-V1} variant exceeds \textit{CodeV-R1}'s performance on VerilogEval-v2, VerilogEval-v2-NTU, and CVDP-cid02\&03, lagging only on RTLLM-v2.

Across the suite, \textit{SiliconMind-V1} models generally outperform \textit{Qwen3-14B} and \textit{Qwen3-32B}~\cite{qwen3} (with only the \textit{Qwen2.5-Coder-7B-Instruct} variant trailing on VerilogEval-v2 and VerilogEval-v2-NTU). A striking example of this efficiency is our \textit{Qwen3-4B-Thinking-2507} variant, which, despite being 171x smaller, nearly matches \textit{DeepSeek-R1-0528} with a performance gap of only 0.8-6.3\% across all benchmarks.

\subsection{Ablation Study} \label{sec:eval_ablation}

Table~\ref{tab:eval_ablation} details the percentage changes in \textit{pass@1} performance from our framework's progression. We see the most profound \textit{pass@1} improvement when progressing from the base model to \textit{SiliconMind-dev} using the \textit{regular} inference strategy, averaging 23.3\% across all base models and benchmarks. Following up is the progression from \textit{SiliconMind-dev} to \textit{SiliconMind-V1} using the \textit{regular} strategy, adding another 4.6\% on average. The \textit{Deep Thinking} and \textit{Agentic} strategies further actualize the \textit{SiliconMind-V1} models' potential, topping \textit{regular} by 0.8\% and 2.7\% respectively on average.

Looking more closely, CVDP-cid02\&03 is the hardest benchmark to improve on. The \textit{Qwen3-4B-Thinking-2507} based \textit{SiliconMind-V1} even experienced a minor performance drop on the benchmark when adopting the \textit{Deep Thinking} strategy. Between base models, the magnitude of overall improvement is inversely proportional to their starting strength (as reported in Table~\ref{tab:eval_general}), thus, \textit{Olmo-3-7B-Think} saw the greatest gains, followed by \textit{Qwen2.5-Coder-7B-Instruct}, \textit{Qwen3-4B-Thinking-2507}, and \textit{Qwen3-8B}. This also confirms that the model's grasp of Verilog design is bottlenecked by the amount of training data.

\subsection{Tailored vs. Non-Tailored Data} \label{sec:eval_tailored}

As mentioned at the end of \sectionname~\ref{sec:method:internalsft}, we developed two testing and debugging curricula during the \textit{Self-Correction} phase: one for the Qwen family and another tailored to the \textit{Olmo-3-7B-Think}-based model. Table~\ref{tab:eval_tailored} compares the $pass@1$ performance between the \textit{Olmo-3-7B-Think} based \textit{SiliconMind-V1} trained on the tailored curriculum versus the one trained on Qwen's, both utilizing the \textit{Deep Thinking} inference strategy. On average, the tailored curriculum increases performance by 1.7\% across all benchmarks, justifying the additional computational resources invested in its creation.

\begin{table}[]
\centering
\caption{Pass@1 Improvement from Tailored Curriculum for \textit{SiliconMind-V1-Olmo-3-7B-Think}.}
\label{tab:eval_tailored}
\begin{tabular}{cccc}
\toprule
\multirow{2.5}{*}{\textbf{Tailored?}} & \multicolumn{3}{c}{\textbf{Pass@1 Performance with Deep Thinking (\%)}} \\ \cmidrule(lr){2-4} 
 & {\textbf{RTLLM-v2}} & {\textbf{VerilogEval-v2-NTU}} & {\textbf{CVDP-cid02\&03}} \\ \midrule
NO  & 57.8\phantom{$_{\textcolor{darkpastelgreen}{\text{\footnotesize+2.0}}}$} & 74.7\phantom{$_{\textcolor{darkpastelgreen}{\text{\footnotesize+0.8}}}$} & 18.2\phantom{$_{\textcolor{darkpastelgreen}{\text{\footnotesize+2.2}}}$} \\
YES & 59.8$_{\textcolor{darkpastelgreen}{\text{\footnotesize+2.0}}}$ & 75.5$_{\textcolor{darkpastelgreen}{\text{\footnotesize+0.8}}}$ & 20.4$_{\textcolor{darkpastelgreen}{\text{\footnotesize+2.2}}}$ \\ \bottomrule
\end{tabular}
\end{table}

\subsection{Cost-Performance of Different Inference Strategies} \label{sec:eval_cp}

\figurename~\ref{fig:eval_costperf} illustrates the cost-performance trade-offs of our inference strategies. In the following analysis, the \textit{regular} inference strategy serves as the baseline. Across all \textit{SiliconMind-V1} variants and benchmarks, the \textit{Deep Thinking} strategy yields a 0.53-1.28\% increase in $pass@1$ performance while scaling the number of response tokens by 1.14--1.26x.

For the \textit{Agentic} strategy, limiting the number of test/debug agent interactions to one increases performance by 2.0\% on average, at the expense of 2.1x more tokens. When allowing up to two or three interactions, the marginal gains in performance and token costs diminish, as the strategy terminates once the model is satisfied with its response. Specifically, the second interaction increases the average performance gain to 2.5\% (at 2.6x cost), while a third interaction reaches 2.8\% (at 2.9x cost).

CVDP-cid02\&03 represents the most difficult benchmark for trading token cost for performance. The \textit{Qwen2.5-Coder-7B-Instruct}-based \textit{SiliconMind-V1} model achieved only a 0.4\% performance increase after the third agentic interaction; notably, stopping at the first or second interaction even resulted in performance regression. Among the \textit{SiliconMind-V1} variants, the \textit{Olmo-3-7B-Think} model exhibited the most favorable cost-performance trade-off.

%% file: paper_body/conclusion.tex
\section{Conclusion} \label{sec:Conclusion}

In this work, we presented a unified framework that combines multi-agent distillation with test-reasoning workflows for Verilog code generation, culminating in the SiliconMind-V1 model series. By automating the creation of reasoning-oriented training data and testbenches through a multi-agent collaboration pipeline, the framework addresses challenges of data scarcity and quality in hardware design. The distilled LLMs are guided by a test-time inference engine to iteratively generate, test, and debug Verilog code without relying on external tools. Comprehensive evaluation demonstrates that our approach significantly outperforms the state-of-the-art, advancing the capabilities of LLM-assisted hardware design.